\documentstyle[12pt,aaspp4]{article}

\begin{document}

\title{The Mass of the Centaurus A Group of Galaxies}
\author{Sidney van den Bergh}
\affil{
Dominion Astrophysical Observatory \\
National Research Council \\ 
5071 West Saanich Road \\
Victoria, British Columbia, V8X 4M6 \\
Canada \\
vdb@dao.nrc.ca }

\begin{abstract}

The mass M, and the radius R$_{\rm h}$, of the Centaurus A group are 
estimated from the positions and radial velocities of 30 probable 
cluster members.  For an assumed distance of 3.9 Mpc it is found 
that R$_{\rm h} \sim 640$~kpc.  The velocity dispersion in the Cen A group is 
$114 \pm 21$ km s$^{-1}$.  From this value, and R$_{\rm h} = 640$~kpc, 
the virial theorem yields a total mass of $1.4 \times 10^{13}$~M$_{\sun}$
for the Cen A group.  
The projected mass method gives a mass of $1.8 \times 10^{13}$~M$_{\sun}$.  
These values suggest that the Cen A group is about seven times as 
massive as the Local Group.  The Cen A mass-to-light ratio is found 
to be M/L$_{\rm B} = 155-200$ in solar units.  
The cluster has a zero-velocity radius R$_0 = 2.3$~Mpc.

\end{abstract}


\section{Introduction}
The loose clustering of galaxies surrounding Centaurus A was called the NGC 5128 
Group by de Vaucouleurs (1975), who designated it G4 because it was the fourth 
closest group of galaxies in his compilation.  This group is primarily of 
interest because its brightest member is the galaxy NGC 5128, which contains 
the radio source Centaurus A.  This object is also [with the exception of 
Maffei 1 (van den Bergh 1971, Huchtmeier, Karachentsev \& Karachentsev 1999)] 
the nearest early-type giant galaxy.  De Vaucouleurs recognized the galaxies 
NGC 4945, 5068, 5102, 5128, 5236 and 5253 
as probable group members.  Using more recent data Hesser et al. (1984) 
assigned 20 galaxies to this group.  From their data these authors 
derived a group mass of $8 - 25 \times ({\rm D}/5 {\rm Mpc}) \times 10^{12}\ {\rm M}_{\sun}$.  
Using \underline{Hubble Space Telescope} (HST) observations of the tip of the red giant 
branch in NGC 5128 Harris, Poole \& Harris (1998) obtained a distance modulus 
(m-M)$_0 = 27.89 \pm 0.15$, corresponding to a distance of $3.9 \pm 0.3$~Mpc. 
This value is in good agreement with (m-M)$_0 = 27.8 \pm 0.2\ (3.6 \pm 0.2\ {\rm Mpc})$ 
that Soria et al. (1996) had previously found using the same method.  Independent 
distance estimates are by Hui et al. (1993) who, after correction to modern Local 
Group distances, found (m-M)$_0 = 27.97 \pm 0.14$ from the luminosity function of 
planetary nebulae, and by Tonry \& Schechter (1990) who obtained (m-M)$_0 = 27.53 \pm 0.25$, 
corresponding to $3.2 \pm 0.4$~Mpc, from surface brightness fluctuations in NGC~5128.

It would be of interest to search for Cepheids among 
the young blue stellar populations (van den Bergh 1976a) 
that are associated with the dark band of gas and dust which 
crosses the face of NGC~5128.  It is generally believed that 
the present, rather chaotic, morphology of NGC~5128 is due 
to the fact that this object recently merged with a spiral 
galaxy.  The S0pec galaxy NGC~5102 in the Cen A group is 
presently undergoing a burst of star formation, which may 
have been triggered by gas captured from the luminous Sc 
spiral NGC~5253 about $4 \times 10^8$ years ago  (van den Bergh 1976b).

An independent check on the distance of the Centaurus A group is 
provided by HST observations of Cepheids in NGC 5253 (Saha et al. 1995), 
which yields (m-M)$_0 \sim 28.1$, corresponding to ${\rm D} = 4.2$~Mpc.  In the subsequent 
discussion a distance of 3.9 Mpc will be adopted for both NGC~5128 and for the 
Cen A cluster.  From a historical point of view it is of interest to note that 
this value is, within its errors, indistinguishable from the distance ${\rm D} = 4.0$~Mpc 
adopted by de Vaucouleurs (1975).  This serves to illustrate the fact that the 
extragalactic distance scale, out to distances of  $\sim 10$~Mpc, has remained essentially 
unchanged for a quarter of a century.

\section{Cluster Distance and Radius}

A catalog of possible members of the Centaurus group has been published 
by C\^{o}t\'{e} et al. (1997).  More recently Banks et al. (1999) have obtained HI 
observations for a number of cluster suspects.  Table 1 gives data on 30 probable 
group members.  Following Banks et al.  the galaxy CEN 5, which has a heliocentric 
radial velocity of only  $+130$~km~s$^{-1}$, has been omitted as a probable foreground object.  
Alternatively it might be a background galaxy that is superposed on a local 
high-velocity cloud.  Also omitted is NGC~5068, which lies 22$^{\circ}$ north of the main 
concentration of Centaurus A group members.  Data listed in the table are:  
(1) Galaxy name, (2) alias, (3) angular distance from NGC  5128, (4) radial velocity 
corrected to the centroid of the Local Group, assuming a solar velocity of 306~km~s$^{-1}$ 
towards $\ell = 99^\circ$, ${\rm b} = -4^\circ$ (Courteau \& van den Bergh  1999), 
and (5) the galactic HI content in units of $10^6\ {\rm M}_{\sun}$ from Banks et al. (1999).

The data plotted by Banks et al. show that the Centaurus group has a rather 
irregular structure.  Nevertheless, it will be assumed that this group is dynamically 
centered on NGC~5128, which is its most luminous (and presumably most massive) member.  
The distribution of the angular distances of individual galaxies in the Centaurus group 
from NGC~5128 is shown in Figure~1. Taken at face value the data plotted in the figure 
suggest that half of the members of the Centaurus group are situated within  $\sim 9.^\circ5$ 
of  NGC~5128.  At a distance of 3.9 Mpc this yields R$_{\rm h} = 640$~kpc.  The true half-mass radius 
is expected to be smaller than this value because a significant fraction of the cluster 
mass is associated with the central galaxy NGC~5128.  On the other hand the incompleteness 
of the sample at R(NGC 5128)$\ > 14^\circ$ will lead to an underestimate of the true half-mass 
radius.  It is noted in passing that the value R$_{\rm h} \sim 640$~kpc for the Cen A group is larger 
than the value R$_{\rm h} \sim 350$~kpc, which Courteau \& van den Bergh (1999) find for the Local Group.  
Furthermore the  $\sim 900$~kpc projected separation, between the dominant galaxies NGC~5128 
and NGC~5253, is slightly larger than the 760 kpc linear separation (van den Bergh 2000) 
of the Milky Way system and the Andromeda galaxy.

\section{Mass of the Centaurus A Cluster}

If it is assumed that the Cen A cluster is in virial equilibrium, and that its velocity 
ellipsoid is isotropic ($\sigma^2 = 3 \sigma_{\rm r}^2$), then its mass can be computed from its velocity 
dispersion and half-mass radius (Spitzer 1969).  According to Binney \& Tremaine (1987, Eq. 4-80b)
\begin{equation}
{\rm M}_{\rm v}  \simeq  \frac{7.5}{\rm G} \langle \sigma_{\rm r}^2 \rangle {\rm r}_{\rm h} = 
1.74 \times 10^6 \langle \sigma_{\rm r}^2 \rangle {\rm r}_{\rm h} {\rm M}_{\sun},
\label{eq:1}
\end{equation}
in which $\sigma_{\rm r}$ is in km~s$^{-1}$ and r$_{\rm h}$ is in kpc.  From the data in 
Table 1 the radial velocity dispersion $\sigma_{\rm r} = 114 \pm 21$~km~s$^{-1}$.  
Substituting this value into Eqn. (1) yields a total cluster mass of 
$1.4 \times 10^{13}\ {\rm M}_{\sun}$.  
For comparison it is noted that Courteau \& van den Bergh (1999) found a mass of only 
$2.3 \times 10^{12}\ {\rm M}_{\sun}$ for the Local Group.  Reasons for this difference 
are that (1) the radial velocity dispersion in the Local Group is only 
$\sigma_{\rm r} = 61$~km~s$^{-1}$, and (2) the adopted half-mass radius of Cen A is 
R$_{\rm h} = 0.64$~Mpc, compared to R$_{\rm h} \simeq 0.35$~Mpc for the Local Group.

	Alternatively the mass of the Centaurus A group may be estimated using the projected 
mass method of Bahcall \& Tremaine (1981).  According to Heisler, Tremaine, \& Bahcall (1985) 
the projected mass estimator may be expressed as
\begin{equation}
{\rm M}_{\rm p} =  \frac{10.2}{\rm G(N - 1.5)} \sum_i {\rm V}^2_{\rm r,i}\, {\rm R}_{\rm i},
\label{eqn:2}
\end{equation}
in which N is the number of cluster galaxies and G is the gravitational constant. From 
the data in Table 1 this yields M$_{\rm p} = 1.8  \times 10^{13}\ {\rm M}_{\sun}$, 
which is in reasonable agreement with the virial mass of 
M$_{\rm v} = 1.4  \times 10^{13}\ {\rm M}_{\sun}$. 
The $14 - 18 \times 10^{12}\ {\rm M}_{\sun}$ mass of the Cen A group is about seven 
times greater than the ($2.3 \pm 0.6) \times 10^{12}\ {\rm M}_{\sun}$ 
mass that Courteau \& van den Bergh (1999) derived for the Local Group.

The mean velocity of the Cen A group, relative to the centroid of the Local Group, 
is  $+278 \pm 21$~km~s$^{-1}$.  This value is, within its errors, indistinguishable 
from the $+295$~km~s$^{-1}$ velocity of NGC~5128 relative to the Local Group.  
The cluster velocity of $+278$~km~s$^{-1}$, in conjunction with a distance of 
3.9~Mpc, yields a Hubble parameter H$_0 = 71$~km~s$^{-1}$~Mpc$^{-1}$.  
The fact that this value is so close to the numerical value of H$_0$, 
that is derived from observations of distant galaxies, suggests that the local 
region of the Universe may be rather cold.

The radius of the zero-velocity surface, which separates the cluster from the Hubble flow 
(Lynden-Bell 1981, Sandage 1986), is given by 

\begin{equation}
{\rm R}_0 [{\rm Mpc}] = \left( \frac{\rm 8GT^2}{\pi^2} {\rm M} \right)^{1/3} 
= 0.154 ({\rm T[Gyr]})^{2/3} ({\rm M[10^{12} M\sun]})^{1/3}.
\label{eq:3}
\end{equation}

Assuming ${\rm M} = 16 \times 10^{12}\ {\rm M}_{\sun}$ and T$ = 14$~Gyr one 
obtains R$_0 = 2.3$~Mpc, corresponding 
to an angular size of 35$^\circ$.  The radius of the zero-velocity surface of the 
Local Group is 1.2~Mpc (Courteau \& van den Bergh 1999).  The distance to the 
Cen A group is 3.9~Mpc, and the radii of the zero-velocity surfaces of the Cen A 
group and Local Group are 2.3~Mpc and 1.2~Mpc, respectively.  The zero velocity 
surfaces of these two clusters therefore almost touch each other.  This suggests 
that the early evolution of the Local Group may have been dynamically affected by the Cen A group.

\section{Luminosity of the Centaurus A Group}
	De Vaucouleurs et al. (1991) list luminosities and absorption values for 16 of the 
brightest galaxies in Table 1.  For an assumed distance of 3.9~Mpc these objects have a 
combined absolute magnitude M$_{\rm B} = -21.9$. Of this luminosity 58\% 
is contributed by NGC~5128 itself.  The three brightest cluster galaxies (NGC~945, NGC~5128 
and NGC~5256) together provide 89\% of the total group luminosity calculated above.  These 
data suggest that the fainter group galaxies provide only a negligible contribution to the 
total group luminosity. Adopting M$_{\rm B} = -21.9$ and M$_{\rm B} (\sun) = +5.48$ 
yields a total cluster luminosity L$_{\rm B} = 9 \times 10^{10}\ {\rm L}_{\rm B} (\sun)$.  
In conjunction with a total cluster mass of $14 - 18 \times 10^{12}\ {\rm M}_{\sun}$, 
this yields M/L$_{\rm B} = 155 - 200$ in solar units.  With 
${\rm h} = ({\rm H}_0/100\ {\rm km\ s}^{-1}\ {\rm Mpc}^{-1}) \sim 0.7$ this is 
quite similar to the value M/L$_{\rm B} \sim 250$~h (in solar units) that Girardi et al. (1999) obtain 
for poor nearby Abel clusters.  However, it is larger than the value
M/L$_{\rm V} = 44 \pm 12$ that Courteau \& van den Bergh (1999) find for the Local Group.

\section{Conclusions}
	The Centaurus A group is an irregular clustering centered on NGC~5128.  This group 
has a mass which is about seven times greater than that of the Local Group.  It is located 
at a distance of  $\sim 3.9$~Mpc, and is found to have a mass-to- light ratio (in solar units) 
of 155--200, which is quite similar to the mass-to-light ratios found for typical poor Abel 
clusters.  The Cen A group has a half-light radius of  $\sim 640$~kpc.  Its zero-velocity 
surface (assumed to be spherical) has a radius R$_0 = 2.3$~Mpc.  Since this value is more 
than half of the distance to NGC~5128, it appears probable that the Cen A cluster may 
have had a significant influence on the early dynamical evolution of the Local Group.  

\acknowledgements
I am indebted to St\'{e}phanie C\^{o}t\'{e} for reading, and commenting on, an earlier version of this paper.

\clearpage
\begin{deluxetable}{llrrr}
\tablecaption{DATA ON CENTAURUS A CLUSTER}
\tablehead{
\colhead{(1)} & 
\colhead{(2)} & 
\colhead{(3)} & 
\colhead{(4)} & 
\colhead{(5)\tablenotemark{a}} \\
\colhead{Galaxy} & 
\colhead{Alias} & 
\colhead{D(5128)} & 
\colhead{V$_{\rm LG}$ (km s$^{-1}$)} & 
\colhead{M$_{\rm HI}$}}
\startdata
ESO 321-G014 & UKS 1243-336 & 14.$^\circ$36 & +333 &     19 \nl

ESO 381-G018 &  & 10.61 & +349 &     13 \nl

ESO 381-G020 &  & 11.98 & +333 &   100 \nl

Cen 8 &  &    4.79 & +350 &     20 \nl

NGC 4945 &  &    7.39 & +299 &   990 \nl

ESO 269-G058 &  &    4.77 & +133 &     21 \nl

1321-31 &  & 11.51 & +355 &     24 \nl

NGC 5102 &  &   6.42 & +225 &   320 \nl

AM 1321-304 &  & 12.05 & +268 &     11 \nl

NGC 5128 & Cen A &   0.00 & +295 &   490 \nl

IC 4247 &  & 12.65 & +191 &     18 \nl

ESO 324-G024 & UKS 1324-412 &   1.58 & +268 &   120 \nl

1328-30 &  & 12.59 &    -28 &     32 \nl

NGC 5206 &  &   5.33 & +312 &    ... \nl

ESO 270-G017 & Fourcade-Fig. &   3.03 & +575 &   740 \nl

NGC 5236 & M83 & 13.31 & +288 & 3700 \nl

1337-39 &  &   3.86 & +246 &     23 \nl

NGC 5237 &  &   2.24 & +111 &     33 \nl

NGC 5253 &  & 11.72 & +175 &   150 \nl

NGC 5264 &  & 13.50 & +281 &    82 \nl

ESO 272-G025 &  & 14.12 & +402 &    ... \nl

ESO 325-G011 & UKS 1342-416 &   3.79 & +306 &    77 \nl
 
ESO 174-G001 &  & 10.98 & +422 &  210 \nl

1348-37 &  &   6.71 & +336 &    12 \nl

ESO 383-G087 & UKS 1346-358 &   8.35 & +100 &    90 \nl

1351-47 &  &   6.02 & +281 &    13 \nl

NGC 5408 &  &   7.20 & +267 &  160 \nl

UKS 1424-460 &  & 11.59 & +150 &    61 \nl

ESO 222-G010 &  & 13.58 & +388 &    26 \nl

ESO 274-G001 &  & 19.53 & +314 &  430 \nl
\enddata
\tablenotetext{a}{These masses by Banks et al. (1999) are based on an assumed 
distance of 3.5 Mpc
and should be multiplied by a factor of 1.24 if the true 
cluster distance is 3.9 Mpc.}
\end{deluxetable}

\clearpage

\begin{figure}[h]
\plotone{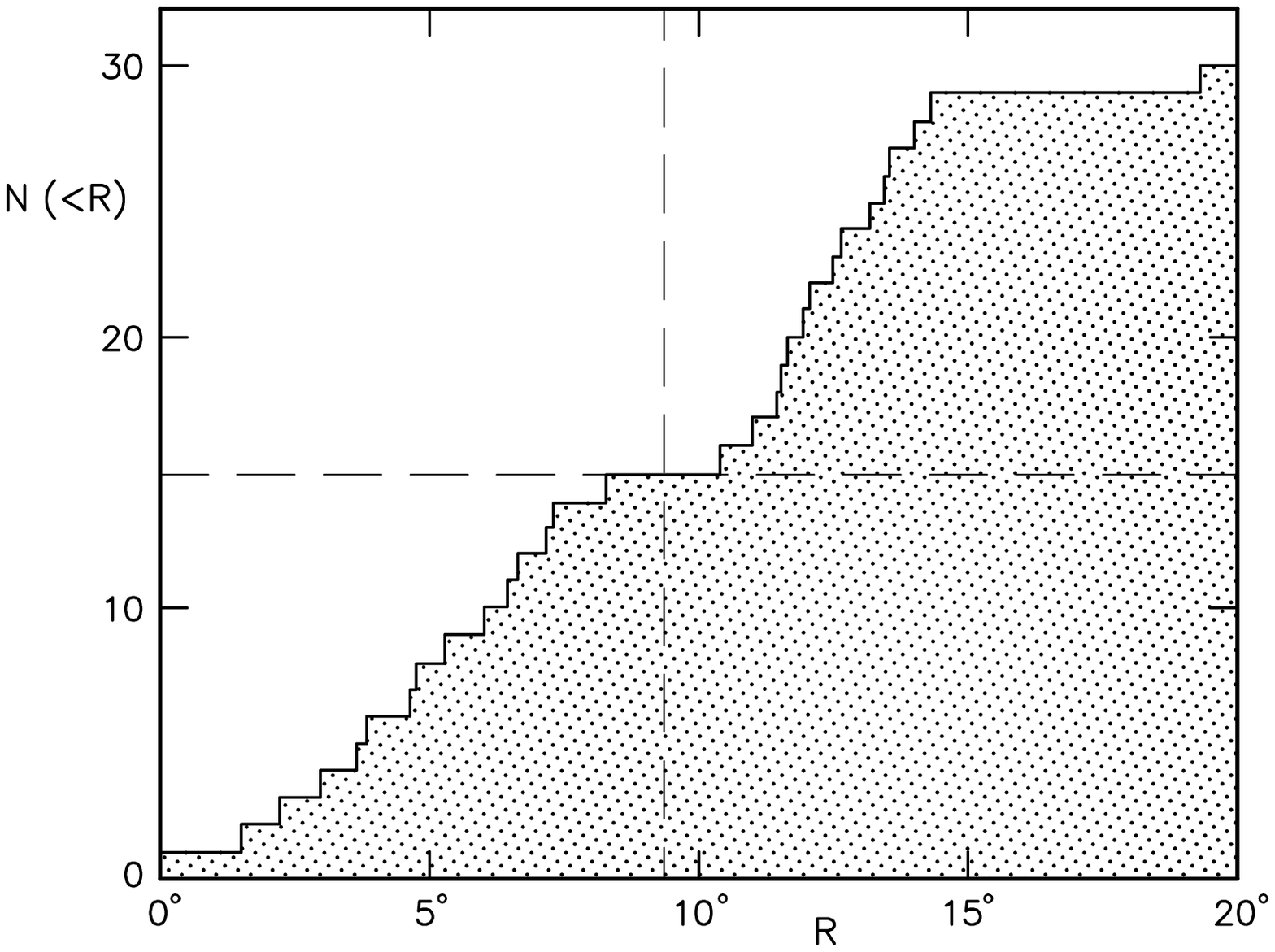}
\caption{Integral distribution of galaxy distances from NGC~5128.  
Half of the known cluster members are located within  $\sim 9.^\circ5$ degrees 
of this galaxy.  The mass distribution in the Cen A group may be more centrally 
concentrated than that of the galaxies.  The data are incomplete for ${\rm R} > 14^\circ$.} 
\label{fig:1}
\end{figure}


\begin{references}
\reference {B81} Bahcall, J.N. \& Tremaine, S. 1981, ApJ, 244, 805
\reference {B99} Banks, G.D. et al. (1999) astro-ph/9906146
\reference {B87} Binney, J. \& Tremaine, S. 1987, Galactic Dynamics, (Princeton:  Princeton 
Univ. Press), p. 214
\reference {C97} C\^{o}t\'{e}, S., Freeman, K.C., Carignan, C. \& Quinn, P.J. 1997, AJ, 114, 1313
\reference {C99} Courteau, S. \& van den Bergh, S. 1999, AJ, in press
\reference {d75} de Vaucouleurs, G. 1975 in Galaxies and the Universe, Eds. A  Sandage, M. 
Sandage, and J. Kristian (Chicago:  Univ. of Chicago Press), p. 557
\reference {d91} de Vaucouleurs, G., de Vaucouleurs, A., Corwin, H.G., Buta, R.J., Paturel, G. \& 
Fouqu\'{e}, P. 1991, (Berlin:  Springer Verlag)
\reference {G99} Girardi, M., Borgani, S., Giuricin, G., Madrossian, F. \& Mezzetti, M. 1999,
Astro-ph/9907266
\reference {H98} Harris, G.L.H., Poole, G.B. \& Harris, W.E. 1998, AJ, 116, 2866
\reference {H85} Heisler, J., Tremaine, S. \& Bahcall, J.N. 1985, 298, 8
\reference {H84} Hesser, J.E., Harris, H.C., van den Bergh, S. \& Harris, G.L.H. 1984, ApJ, 276, 
491
\reference {H99} Huchtmeier, W.K., Karachentsev, I.D. \& Karachentsev, V.E. 1999, astro-ph/9907344
\reference {H93} Hui, X., Ford, H.C., Ciardullo, R. \& Jacoby, G.H. 1993, ApJ, 414, 463
\reference {L81} Lynden-Bell, D. 1981, Observatory, 101, 111
\reference {S95} Saha, A., Sandage, A., Labhardt, L., Schwengeler, H., Tammann, G.A., Panagia, 
N. \& Macchetto, F.D. 1995, ApJ, 438, 8
\reference {S86} Sandage, A. 1986, ApJ, 307, 1
\reference {S96} Soria, R. et al. 1996, ApJ, 465, 79
\reference {S69} Spitzer, L. 1969, ApJL, 158, 139
\reference {T90} Tonry, J.L. \& Schechter, P.L. 1990, AJ, 100, 1794
\reference {v71} van den Bergh, S. 1971, Nature, 231, 35, 1971
\reference {v76} van den Bergh, S. 1976a, ApJ, 208, 673
\reference {v76b} van den Bergh, S. 1976b, AJ, 81, 795
\reference {v00} van den Bergh, S. 2000, The Galaxies of the Local Group, (Cambridge:
Cambridge Univ.Press), in press
\end{references}
\end{document}